%% file: main.tex
\documentclass[letterpaper,twocolumn,10pt]{article}
\usepackage{usenix-2020-09}
\usepackage{tikz}
\usepackage{amsmath}
\usepackage{filecontents}

\newcommand{\proj}{\textsc{FlexInfer}}
\input{tex/preamble}

\date{}

\begin{document}

\title{vTensor: Flexible Virtual Tensor Management for Efficient LLM Serving}

\author{
Jiale Xu$^{\text{1,2,*}}$,\enskip
Rui Zhang$^{\text{3,*}}$, \enskip
Cong Guo$^{\text{1,2,}}$\textsuperscript{\S},\enskip
Weiming Hu$^{\text{1,2}}$, \enskip
\\
Zihan Liu$^{\text{1,2}}$,  \enskip
Feiyang Wu$^{\text{1,2}}$, \enskip
Yu Feng$^{\text{1,2}}$, \enskip
Shixuan Sun$^{\text{1,2}}$, \enskip
Changxu Shao$^{\text{3}}$, \enskip
\\
Yuhong Guo$^{\text{3}}$,\enskip
Junping Zhao$^{\text{3},}$\textsuperscript{\S},\enskip
Ke Zhang$^{\text{3}}$,\enskip
Minyi Guo$^{\text{1,2}}$,\enskip
Jingwen Leng$^{\text{1,2}}$\enskip
\\
{$^{\text{1}}$Shanghai Jiao Tong University,\enskip $^{\text{2}}$Shanghai Qi Zhi Institute,\enskip $^{\text{3}}$Ant Group\enskip}
\newthanks{Equal contribution.} 
\newthanks{Cong Guo and Junping Zhao are corresponding authors.}
}

\maketitle

\input{tex/abstract}

\pagestyle{plain}

\input{tex/introduction}

\input{tex/background}

\input{tex/motivation}
\input{tex/overview}
\input{tex/vtensor}

\input{tex/felxinfer}

\input{tex/evaluation}

\input{tex/relatedwork}
\input{tex/conclusion}

\bibliographystyle{plain}
\bibliography{references}

\end{document}

%% file: tex/preamble.tex
\usepackage{lipsum}
\usepackage[normalem]{ulem}
\usepackage{tikz}
\usepackage{graphicx}
\usepackage{subcaption}
\usepackage{caption}
\usepackage{algorithmic}
\usepackage[ruled,vlined,linesnumbered,algo2e]{algorithm2e}
\usepackage{amsmath}
\usepackage{ulem}
\usepackage{hyperref}
\usepackage{enumitem}
\usepackage{marvosym}
\usepackage{fontawesome}

\newcommand{\Fig}[1]{Figure~\ref{#1}}

\newcommand{\Tbl}[1]{Table~\ref{#1}}
\newcommand{\Sec}[1]{Section~\ref{#1}}

\newcommand\newthanks[1]{%
  \begingroup
  \renewcommand\thefootnote{}\thanks{#1}%
  \addtocounter{footnote}{2}%
  \endgroup
}

%% file: tex/abstract.tex
\begin{abstract}
Large Language Models (LLMs) are widely used across various domains, processing millions of daily requests. 
This surge in demand poses significant challenges in optimizing throughput and latency while keeping costs manageable. 
The Key-Value (KV) cache, a standard method for retaining previous computations, makes LLM inference highly bounded by memory. 
While batching strategies can enhance performance, they frequently lead to significant memory fragmentation.
Even though cutting-edge systems like vLLM mitigate KV cache fragmentation using paged Attention mechanisms, they still suffer from inefficient memory and computational operations due to the tightly coupled page management and computation kernels.

This study introduces the vTensor, an innovative tensor structure for LLM inference based on GPU virtual memory management (VMM). 
vTensor addresses existing limitations by decoupling computation from memory defragmentation and offering dynamic extensibility. 
Our framework employs a CPU-GPU heterogeneous approach, ensuring efficient, fragmentation-free memory management while accommodating various computation kernels across different LLM architectures. 
Experimental results indicate that vTensor achieves an average speedup of 1.86$\times$ across different models, with up to 2.42$\times$ in multi-turn chat scenarios. 
Additionally, vTensor provides average speedups of 2.12$\times$ and 3.15$\times$ in kernel evaluation, reaching up to 3.92$\times$ and 3.27$\times$ compared to SGLang Triton prefix-prefilling kernels and vLLM paged Attention kernel, respectively. 
Furthermore, it frees approximately 71.25\% (57GB) of memory on the NVIDIA A100 GPU compared to vLLM, enabling more memory-intensive workloads.

\end{abstract}

%% file: tex/introduction.tex
\section{Introduction}\label{sec:introduction}

The rapid advancement of large language models (LLMs)\cite{zhang2022opt,brown2020language} has transformed various domains, including natural language processing\cite{min2023recent}, recommendation systems~\cite{cui2022m6rec}, and autonomous driving~\cite{cui2023survey}.
Prominent models such as ChatGPT~\cite{GPT3,GPT4}, LLaMA~\cite{touvron2023llama,touvron2023llama2}, and Mixtral~\cite{jiang2024mixtral} now process millions of requests daily.
This surge in demand challenges the underlying infrastructure, primarily due to the high computational costs of processing these requests.
Therefore, serving systems must balance high throughput with low latency to ensure cost-effectiveness and maintain user satisfaction.

In conjunction with the autoregressive mechanism of LLMs~\cite{vaswani2023attention}, the Key-Value (KV) cache~\cite{pope2022efficiently} has become a standard method for storing previous Keys (K) and Values (V) without recomputation.
This approach computes Attention only to the newest tokens, thereby accelerating LLM execution.
As a result, LLMs have become highly memory-bound applications, as frequently noted in the literature~\cite{kwon2023efficient,prabhu2024vattention}.
To enhance computational efficiency, batching multiple requests into a single batch is a widely adopted optimization strategy to improve LLM serving performance~\cite{280922,patel2024splitwise}.

\begin{figure}[t]
    \vspace*{3mm}
    \begin{center}
    \includegraphics[width=0.47\textwidth]{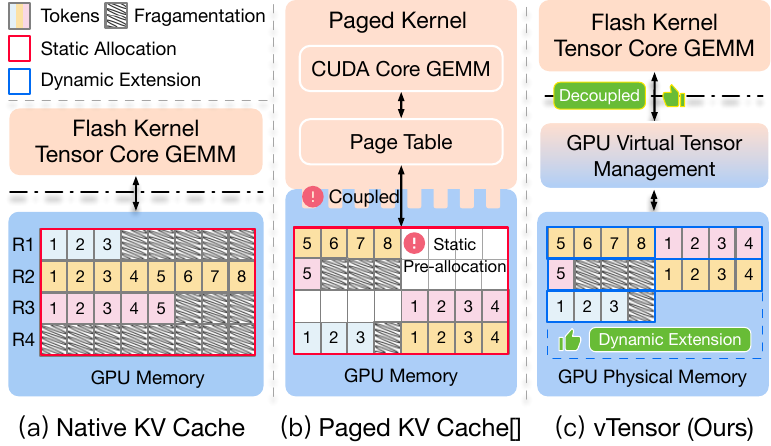}
    \caption{Three KV cache memory management strategies: (a) Native KV cache with native GPU allocation has a large volume of fragmentation; (b) vLLM adopted paged memory management to eliminate most fragmentation but with tightly coupled kernels; and (c) vTensor decouples computation and memory allocation with more flexible management.}
    \label{fig:intro}
    \end{center}
\end{figure}

However, batching often leads to significant memory fragmentation within the KV cache owing to different sequence lengths of different requests, as illustrated in \Fig{fig:intro}(a).
Traditional LLM systems typically pre-allocate contiguous memory space for up to 4096 tokens, based on a predefined maximum request length.
To ensure compatibility with optimized libraries such as cuBLAS~\cite{cuBLAS,torres2024evaluation} and efficient execution, all requests are padded to uniform length and batch size, resulting in substantial fragmentation.
According to previous studies~\cite{kwon2023efficient} and our observations, the fragmentation ratio can reach up to approximately 70\% of the total 80~GB memory in the GPU A100, causing significant resource wastage.

Therefore, defragmenting the KV cache is critical for performance improvements as it reduces memory usage and increases the inference batch size.
Previous studies, such as vLLM~\cite{vLLM}, introduced paged Attention, a strategy for managing KV cache via a page table, a classic defragmentation mechanism.
They implemented a page table and an address translator within a custom CUDA kernel for GPUs, while continuing to use conventional GPU memory allocation methods, as illustrated in \Fig{fig:intro}(b).
By offering a larger batching space, paged Attention can outperform traditional KV cache management in the overall system performance.

Unfortunately, this paged Attention mechanism is tightly coupled with computation kernels, imposing significant constraints on memory and computation flexibility.
First, since paged Attention statically pre-allocates the KV cache with a specific paged data structure, its occupied memory cannot be dynamically re-allocated to other data structures including activations and additional LLM inference instances, which occupies about 83.2\% of GPU memory, leading to substantial idle capacity on the GPU device.
Second, paged Attention introduces two main drawbacks to computational efficiency due to the coupled kernel: (1) the coupled kernel significantly hampers the overall performance of the LLM; (2) the intricate coordination required between page management and computation kernels leads to extremely high development costs for new LLM features.
Our experiments indicate that the potential performance loss is up to 59\% slowdown for emerging LLM applications such as single-generation, multi-turn chat, and prefix sharing, due to the limitations imposed by paged Attention and attentions that are modified to support paged KV cache.

To address these challenges, we introduce a novel LLM inference framework called \textbf{FlexInfer}.
FlexInfer is based on the \textbf{vTensor} (virtual tensor), an efficient abstraction for managing GPU's memory.
As shown in \Fig{fig:intro}(c), vTensor offers two main advantages: computation decoupling and dynamic memory extensibility.
The core design principle of vTensor is to separate memory defragmentation from the original computation kernel, thereby inherently enhancing memory and computational flexibility in LLM serving systems.

\paragraph{Computation Flexibility}
According to our observations, coupling computation with the page table is not necessary for defragmenting the KV cache.
We propose a more efficient method to decouple computation from memory defragmentation kernels.
Firstly, vTensor offers an application interface (API) nearly identical to that of traditional KV cache management, without impacting the computation kernel.
From the viewpoint of computation kernels, vTensor remains transparent and behaves like a standard CUDA-allocated tensor.
This benefit stems from vTensor's robust abstraction and effective management of GPU virtual memory.
We have developed a series of {operators} to manage all memory-related {actions} in the KV cache, including allocation, extension, deallocation, and others.
These {operators} can be effectively implemented via GPU's low-level virtual memory management (VMM) APIs~\cite{nvidia2024cuda,nvidia2024vm}.
To enhance execution efficiency, we integrate vTensor within the LLM serving system to create the FlexInfer framework, enabling the scheduling of memory-related operations without disrupting the original computation.

\paragraph{Memory Flexibility}
Based on the abstraction provided by vTensor, we can readily implement a memory defragmentation strategy.
Unlike paged Attention, which relies on CUDA kernel execution on the GPU, vTensor operates its memory management on the CPU using C/C++ implementation.
vTensor maintains a page table for virtual memory management and tracks unallocated memory.
When a GPU memory allocation or deallocation request is submitted to vTensor, it promptly resolves the request, devises an optimized strategy, and offloads the solution to the GPU for allocating physical memory chunks.
Due to the regularity of LLM inference, we can pre-allocate memory for new tokens for the subsequent iteration in advance, highlighting vTensor's key benefit: dynamic memory extendability for LLM's auto-regressive processing.
This approach lets us manage all memory scheduling using CPU-based concurrent management of GPU virtual memory, which offers greater flexibility.
As a result, all memory management operations can be efficiently overlapped with and hidden by LLM computations, ensuring no overhead impacts on LLM serving on FlexInfer.

Finally, \textbf{FlexInfer} employs a {CPU and GPU heterogeneous framework}, providing efficient and fragmentation-free memory management based on \textbf{vTensor} abstraction and supporting diverse computation kernels across various LLM architectures.
FlexInfer achieves greater efficiency than previous paged KV cache systems while maintaining more flexible defragmentation capabilities.
Our experiments demonstrate that FlexInfer can achieve an average of 1.86$\times$ speedups across 3 applications on 3 predominant LLMs and up to 3.92$\times$ and 3.27$\times$ performance improvements for \proj{} kernels compared to SGLang triton prefix prefilling kernels and vLLM paged Attention kernels, respectively.
Moreover, FlexInfer offers more flexible memory management for the KV cache and frees about 71.25\% (57GB) of memory on the A100 GPU, facilitating more memory-intensive scheduling.
Overall, this work makes the following contributions:
    \begin{itemize}[leftmargin=*]
\item We identify and evaluate the inefficiency of existing paged KV cache memory management in LLM serving systems and their impact on performance.
\item We introduce and implement vTensor, a new virtual memory abstraction that offers a continuous virtual address and a suite of primitives for managing KV cache memory using CUDA virtual memory management controls.
\item We design an efficient vTensor-based inference framework, \proj{}, which enhances both computational and memory flexibility for LLM serving systems.
\item We evaluate \proj{} across various LLM serving applications and demonstrate its superior performance compared to existing paged KV cache systems, achieving improvements up to 2.4 $\times$ in end-to-end evaluation and up to 3.92 $\times$ in computation kernels evaluation with equivalent levels of memory defragmentation.
    \end{itemize}

%% file: tex/background.tex
\section{Background}\label{sec:background}


\subsection{LLM Inference}
\label{sec:llm-intro}

Generative large language models (LLMs)~\cite{zhang2022opt,touvron2023llama,touvron2023llama2,brown2020language} operate in an auto-regressive mode, producing tokens sequentially.
LLMs typically perform two fundamental operations: linear projection operation and attention operation.
The linear projection operation first calculates query ($Q$), key ($K$), and value ($V$) tensors using their respective weights ($W_Q$, $W_K$, and $W_V$).
Subsequently, the attention mechanism computes the similarity between $Q$ and $K$, applying this similarity to weight the corresponding $V$ values and compute the attention output.
Finally, the linear projection operation, known as the feed-forward network, processes the attention output.

%

The user-facing inference process of generative LLMs typically consists of two distinct stages: the prefill stage and the decode stage.~\cite{zhong2024distserve}
In the prefill stage, the model receives a prompt sequence that sets the context for subsequent text generation and produces the initial token.
In addition, the prefill stage cached the $K$ and $V$ tensors to prevent redundant recomputation in the subsequent decode stage.



Following the prefill stage, the decode stage uses the last token and the KV cache to iteratively generate the next token and update the KV cache.
In other words, the KV cache continues to expand until the inference process concludes.



\subsection{LLM Serving Optimizations}
This section delves into recent LLM serving advancements, such as continuous batching and prefix caching, which are critical for maximizing serving efficiency.

\subsubsection{Continuous Batching}
 

Continuous batching~\cite{280922,agrawal2024taming} is a dynamic strategy that replaces a completed request in a batch with a new one immediately after completion.
It applies iteration-level scheduling by determining the batch size per iteration rather than waiting for all requests to complete.
This method significantly boosts GPU utilization by allowing the insertion of a new request as soon as another completes.



\subsubsection{Prefix Cache}

In typical LLM-based applications, the system prompt often remains constant across various requests.
In multi-turn dialogues like ChatGPT~\cite{GPT3,GPT4}, each new message builds on the context of all preceding messages.
Both scenarios can lead to the repeated computation of the key-value (KV) cache for the same prompts, which proves to be computationally costly and time-consuming.



Accordingly, prior studies suggest an optimization technique known as prefix caching~\cite{vllm2023apc}, which stores the KV cache for shared prompt prefixes to eliminate redundant computations and reduce the initial token generation latency.
When a new request arrives, the LLM can bypass recomputing the KV cache for the cached parts, thereby reducing initial token computation time. This is especially beneficial in applications with long and repeated prompts or frequent interactions, such as chatbots~\cite{vicuna2023,mekić2024enhancing} or virtual assistants~\cite{bai2022training,fei2024empathyear}, where the same context is repeatedly processed.


The prefix caching improves the performance of LLM-based applications.
However, it also increases GPU memory usage and poses challenges in combining prefix cache optimizations with efficient computation kernels.


\subsubsection{KV Cache Optimization: GQA and MQA}


KV cache in attention operation is a key problem for LLM inference with the memory-bound scenario.
Multi-head attention (MHA) involves each query head corresponding to a distinct key and value head, resulting in significant KV cache memory consumption.
To address this, several KV cache optimization technologies have been developed.
Multi-query attention (MQA)~\cite{shazeer2019fast}, all query heads share a single key and value heads.
Grouped-query attention (GQA)~\cite{ainslie2023gqa} dividing query heads into $G$ groups, where each group shares a single key head and value head.


\subsection{LLM Serving Systems}

LLM serving systems such as vLLM, TensorRT-LLM ~\cite{TensorRT}, LMDeploy~\cite{LMDeploy} and TGI~\cite{TGI}, is a framework that supports the serving process of LLMs. 
Users can deploy their LLMs and process serving requests or send requests to the system via the network for inference feedback. 

Existing LLM serving systems have integrated a stack of optimizations, including the aforementioned ones, and paged KV-cache, to improve the efficiency of LLM inference. 
However, most of these optimizations are memory-related, trading more memory for less computation or reducing the memory footprint of the KV cache to improve the request batch size and increase the throughput. 
Few systems have focused on the computation efficiency of LLM serving. 

One obvious reason is that the existing paged KV-cache mechanism couples the memory management with the computation kernels. 
This coupling demands massive efforts to migrate existing efficient implementations to the system that supports paged KV-cache, in other words, lacks flexibility.



%

%

\subsection{GPU Virtual Memory Management}
Owing to the increasing need for applications to manage memory with negligible latency and high efficiency, CUDA introduced low-level virtual memory management (VMM)~\cite{nvidia2024cuda,nvidia2024vm,guo2024gmlake}. 
VMM provides primitive operations such as reserve and map to manipulate the virtual address space, offering a finer granularity than traditional methods like \texttt{cudaMalloc}.
The new APIs, including \texttt{cuMemCreate}, \texttt{cuMemAddressReserve}, \texttt{cuMemMap}, and \texttt{cuMemSetAccess}, enable the creation of more efficient dynamic data structures and provide better control over GPU memory usage. 
VMM reduces internal fragmentation and eliminates costly memory operations.
Consequently, VMM delivers significant performance benefits and supports advanced use cases, such as scalable peer mappings and integration with graphics APIs.

%% file: tex/motivation.tex
\begin{figure}[t]
    \begin{center}
    \includegraphics[width=0.47\textwidth]{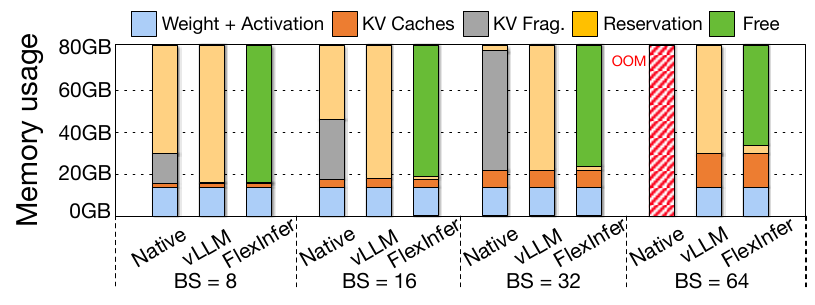}
    \caption{GPU memory usage breakdown using FlashAttention~\cite{FlashAttention} (Native), vLLM~\cite{vLLM}, and FlexInfer on GPU A100 with 80~GB memory. 
    }
    \label{fig:kv_w_ratio}
    \end{center}
    \vspace{-5mm}
\end{figure}

\section{Motivation}\label{sec:motivation}
In this section, we first present the memory issues in the KV cache of prior LLM systems.
Then, we further explore the computation limitations with ill-suited memory defragmentation methods.

\subsection{Memory Flexibility}


To further explore the existing limitations, we conduct a preliminary memory footprint analysis for FlashAttention~\cite{FlashAttention} without memory optimization (native) and vLLM~\cite{vLLM}.
We also compare them with our design, FlexInfer.
All three methods have the same usage in weight, activation, and KV cache since they use the same model and dataset.
As shown in \Fig{fig:kv_w_ratio}, the native method suffers from significant fragmentation, which increases with the batch size (BS) and leads to out-of-memory (OOM) errors when BS is 64.
vLLM attempts to mitigate this by converting KV fragmentation into reserved memory (yellow bar in \Fig{fig:kv_w_ratio}).
vLLM requires static pre-allocation of all available GPU device memory to construct a page table capable of addressing all KV tokens. 
Once the page table is established, the pre-allocated memory is reserved exclusively for the KV cache and cannot be utilized for other allocations, including activations and additional LLM inference instances. 
While vLLM addresses fragmentation issues within a single LLM serving instance, this memory limitation still significantly restricts the overall functionality of the LLM serving system, especially in multi-instance scenarios~\cite{sun2024llumnix}.

Our solution, FlexInfer, introduces two \textbf{key flexibilities}. 
The \textbf{first flexibility} is free memory. 
FlexInfer can dynamically release unused memory, resulting in an average release of {57 GB} of memory across different batch sizes, which is {71.25\%} of the total 80 GB memory on an NVIDIA A100 GPU, as shown in \Fig{fig:kv_w_ratio}.
The \textbf{second flexibility} is the dynamic extension.
Although our approach requires some reservation overhead, approximately 4.99\%, when the batch size (BS) is 64, this is necessary for storing the page table and supporting the dynamic extension.
In addition, this reservation is released immediately when the request ends.

\subsection{Computation Flexibility}
\begin{figure}[t]
    \begin{center}
    \includegraphics[width=0.47\textwidth]{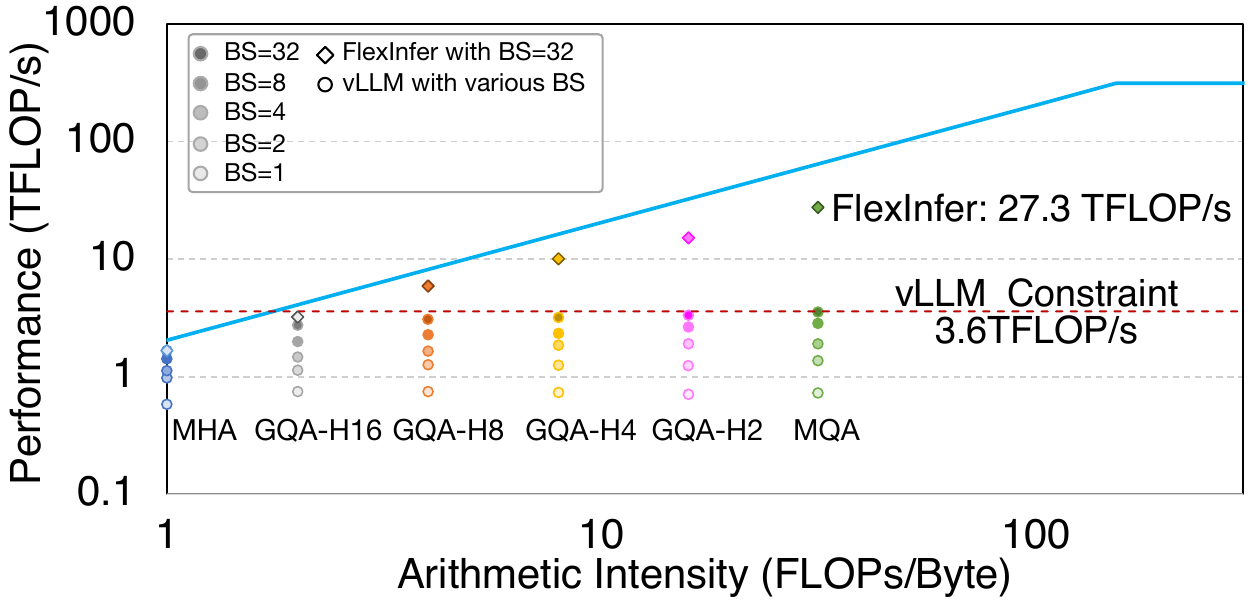}
    \caption{Roofline Model for LLM Attention on GPU A100.
    }
    \label{fig:intro-2}
    \end{center}
    \vspace*{-6mm}
\end{figure}

We also find some serious performance issues in existing systems. 
The native system suffers from fragmentation and inefficiency.
It experiences substantial overhead due to KV cache fragmentation.
vLLM has better overall performance and is now the de facto standard for LLM serving with KV cache defragmentation. 
However, it still has low computational efficiency. 
The paged Attention incurs CUDA kernel overheads associated with address translation, which cannot be overlapped or optimized. 
Moreover, the deeper reason for vLLM is that it significantly limits the programming model. 
The current implementation of vLLM\footnote[1]{In this study, we use vLLM with git commit 973617ae02a4e8e6190674.} relies solely on CUDA cores, which significantly restricts its computational capabilities.
Specifically, vLLM can only use CUDA cores for running paged Attention during the decoding phase, resulting in lower efficiency due to the complex page table translation not being supported by GPU tensor cores. 

\Fig{fig:intro-2} shows a roofline model comparing vLLM and FlexInfer.
The performance of vLLM increases with the increase in batch size.
vLLM and FlexInfer show similar performance in the MHA with 0.99 arithmetic intensity, which is an extremely memory-bound application, and the CUDA core-only is enough to satisfy the computation requirements.
However, the arithmetic intensity increases from MHA to GQA with decreasing sharing heads and then MQA, which means they have a higher computation/memory ratio and higher computation power requirements.
As such, from GQA-H16 (with 16 sharing heads) to GQA-H2 and MQA, vLLM remains nearly the same performance, finally constrained at 3.6 TFLOP/s, due to the CUDA-kernel-based paged mechanism.
For the LLMs with GQA or MQA, the existing paged KV cache management either uses inefficient CUDA cores or requires extensive effort to optimize kernels, but still with low efficiency.
In contrast, FlexInfer's performance improves with increasing arithmetic intensity.
FlexInfer can easily integrate with efficient CUDA kernels, significantly improving the performance based on computation with GPU tensor core and memory management with CPU.
For LLMs with MQA, FlexInfer achieves 27.3 TFLOP/s, which is \textbf{7.58 $\times$} greater than vLLM.

The existing system exhibits two primary drawbacks in terms of computational efficiency.
Firstly, while vLLM can outperform in certain scenarios for original Attention kernels due to their lower arithmetic intensity, this advantage becomes a significant bottleneck when applying new memory-efficient optimizations. 
This limitation prevents vLLM from achieving further performance gains.
Secondly, the handcrafted kernels in vLLM incur substantial development costs when implementing new LLM features. 
Our observations indicate that developing a new paged Attention mechanism typically requires several weeks to months to support emerging features for rapidly evolving LLM applications and scenarios, such as prefill KV cache \cite{pope2022efficiently}. 
This prolonged development cycle significantly constrains the adaptability of the LLM inference system.

Our solution, FlexInfer, effectively addresses these limitations.
First, the decoupling mechanism ensures we can avoid the previous performance issues. 
Second, the VMM-based KV memory management and its serving support significantly improve computation and memory flexibility.
Next, we will present our detailed design and evaluation of FlexInfer in the following sections.

%% file: tex/overview.tex
\section{Overview}

\begin{figure}[t]
    \begin{center}
    \includegraphics[width=0.475\textwidth]{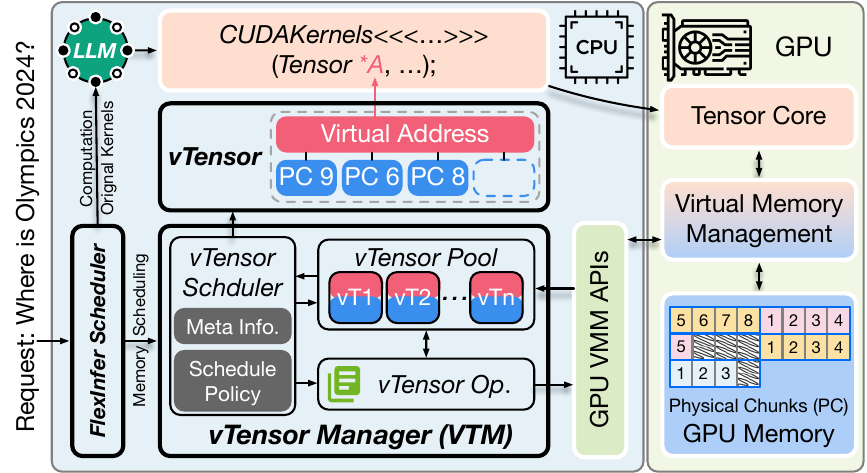}
    \caption{Overview of FlexInfer serving framework.}
    \label{fig:overview}
    \end{center}
    \vspace*{-5mm}
\end{figure}

In this section, we present an overview of our design, which optimizes computation and memory management for LLM serving.
As \Fig{fig:overview} shows, our system consists of two main components: FlexInfer Scheduler and vTensor Manager.

\paragraph{FlexInfer Scheduler} 
As discussed in the previous section, memory defragmentation has become increasingly critical for achieving high performance in LLM serving. 
However, most previous works have focused less on memory-related operations, often coupling them with computation kernels, leading to significant performance loss. 
To address this, we propose FlexInfer, a \textbf{CPU and GPU heterogeneous} framework that decouples and offloads most memory operations to the CPU and hides them by overlapping GPU computation.
Compared to previous works, which run the memory operations (e.g., page table mapping and translating) on GPU~\cite{kwon2023efficient}, the CPU is better at memory-related actions.
When requests are sent to FlexInfer, it decouples the memory and computation operations based on the request's configurations, such as batch size and sequence length. 
For computation, the FlexInfer scheduler employs original highly-optimized kernels to run LLMs on GPU tensor cores, maintaining high computational flexibility and efficiency without arithmetic intensity constraints. 
For memory management, FlexInfer focuses on scheduling memory-related behaviors across different stages of LLM serving. 
Specifically, FlexInfer provides tailored scheduling plans to overlap and hide memory allocation and deallocation with computation during the beginning, prefill, decoding, and end phases.
This strategy significantly mitigates the memory operation overhead in the LLM serving system.

\paragraph{vTensor Manager}
The vTensor Manager (VTM) is the core design for memory defragmentation. It comprises three components: the vTensor Pool (VTP), vTensor Operation (VTO), and vTensor Scheduler (VTS).
The vTensor Scheduler is the key component in the vTensor Manager.
VTS receives instructions from the FlexInfer scheduler and then creates specific policies for each instruction based on meta information. 
This meta information includes multiple data structures (e.g., hash table) to record and track memory-related details, such as unallocated physical chunks (PCs) and vTensor states.
Based on these policies, VTS dispatches specific operations to allocate or deallocate memory through the vTensor Operation (VTO). 
VTO acts as a ``translator'' between the scheduling policies and CUDA low-level VMM APIs. 
It includes all vTensor memory-related actions, translates the policies into primitive APIs, and executes them on the GPU asynchronously.
The execution results are then returned to the vTensor Pool (VTP), which stores all virtual tensors, including virtual memory address mapping information between physical chunks and virtual memory.
VTP also updates the meta information in VTS, which then generates the vTensor for the CUDA kernel or releases a vTensor.

In the following sections, we will introduce the detailed design of FlexInfer and vTensor. We first present the vTensor design in \Sec{sec:vTM} and then illustrate how FlexInfer utilizes and schedules LLM serving based on vTensor in \Sec{sec:flex-sche}.

%% file: tex/vTensor.tex
\section{vTensor and Management}
\label{sec:vTM}
In this section, we first present the vTensor and its management design. 
We will explain how the vTensor manager manipulates and generates a vTensor.

\subsection{vTensor}

\begin{figure}[t]
    \begin{center}
    \includegraphics[width=0.48\textwidth]{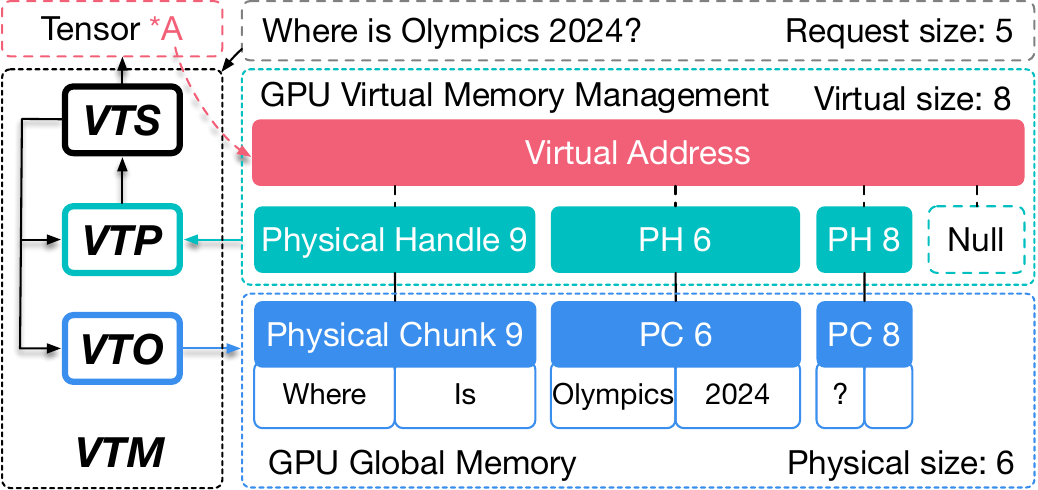}
    \caption{The design of vTensor.}
    \label{fig:vTensor}
    \end{center}
    \vspace*{-3mm}
\end{figure}

vTensor is an efficient abstraction of virtual memory designed for CUDA kernels.
From the perspective of the CUDA kernel, vTensor is a simple tensor pointer to an array, which is identical to a standard CUDA-allocated pointer.
However, vTensor is much more complex than its simple abstraction.

As shown in \Fig{fig:vTensor}, the vTensor pointer is generated by vTensor Manager (VTM).
When a request is sent to vTensor Scheduler (VTS), VTS will create the allocation strategy and then let vTensor Operation (VTO) allocate virtual memory and corresponding physical chunks.
The vTensor pointer ``$*A$'' points to a GPU virtual address (VA), which must be contiguous to be compatible with a standard CUDA-allocated tensor.
Beneath the virtual address, the GPU virtual memory management (VMM) maintains complete mapping information of physical chunks (PCs) registered by the VTM before the CUDA kernel is dispatched to the GPU. 
VMM allows the GPU tensor core to access needed data in the GPU global memory via the virtual address.
Please note that the physical chunks are allocated in the GPU memory, but the physical chunk handles (PHs) and the virtual memory can be accessed and managed by the CPU.
The handles and virtual memory only have the indexing information of the physical chunks without device-host memory transfer.
Specifically, we adopt each physical chunk with a size of 2 MB and store it in GPU global memory. 
Its corresponding handle is only a few bytes in size and is recorded by the vTensor Pool (VTP) and stored in the CPU's main memory.
As such, we have an opportunity to develop a series of novel vTensor-based methods to manipulate the complex defragmentation for the KV cache, which is our basis for vTensor and FlexInfer design.

A key feature of vTensor is the flexible mapping relations among physical chunks,  virtual addresses, and requests.
There are some basic features for the vTensor mappings:
(1) A request can be placed in multiple non-contiguous physical chunks (e.g., PC 9, PC 6, and PC 8 shown in \Fig{fig:vTensor}), but the virtual address must be contiguous;
(2) A physical chunk can be referenced by multiple virtual addresses;
(3) A virtual address may have a larger memory capacity (e.g., 8) than the real physical chunks (e.g., 6), meaning some parts of this virtual address are not mapped to physical chunks.
Therefore, VTM can set the virtual memory address of vTensor with the maximum sequence length but not allocate real physical chunks for the new request, naturally eliminating the fragmentation in the KV cache.
These versatile mapping capabilities of vTensor provide a more flexible and efficient defragmentation solution for the LLM KV cache.

\subsection{vTensor Pool}
\label{sec:vTensor-pool}


The vTensor Pool contains the basic data structure for efficiently managing vTensor for the KV cache.
There are two main types of data structure: ordered set and radix tree (i.e., prefix tree).
As shown in \Fig{fig:vTensor}, two types of information of vTensor are returned to VTP, i.e., a virtual address and corresponding a set of physical handles.
Because of vTensor's flexible mapping mechanism, we store the virtual address and physical handle separately in two order sets, noted vSet and pSet, respectively.
We also build a radix tree, rTree, to address the multi-turn dialogues scenario.

\paragraph{vSet} The vSet consists of a sorted set to store the virtual address of the vTensor.
The virtual address contains a page table that records all pages' corresponding physical handles.
Therefore, The virtual address can be viewed as the vTensor.
As shown in \Fig{fig:vTensor}, we can over-allocate the virtual address size but only map a few pages to real physical handles.
In practice, when a request comes in, we assign a virtual address with a maximum sequence length (e.g., 4096 tokens) but only allocate a few physical chunks (e.g., 256 tokens).

\paragraph{pSet} The pSet is also based on a sorted set to store the physical handles, which has a strict one-to-one correspondence to the physical chunks. 
The pSet maintains the necessary attributes for each physical handle to record its statuses, such as the active state and the vTensor it maps to.
Because multiple vTensors can reference the same physical handle, we manage the physical chunk like the ``hard link'' in a Linux system.  
We assign a reference counter to each physical handle.
Once a physical chunk is allocated on the GPU, its corresponding physical handle is inserted into pSet. 
When the reference counter is 0, the corresponding physical handle is removed from the pSet. 

\paragraph{rTree.} The rTree represents a radix (prefix) tree data structure storing vTensor as the tree node and facilitates prefix matching between different LLM serving requests. 
Once the previous request is finished and a complete vTensor is generated, this vTensor will be a prefix KV cache for the next request. 
Then, the vTensor's pointer in vSet will be inserted into the rTree based on the request's prefix pattern.
When the next request needs its prefix KV cache, the rTree can efficiently search for the prefix vTensor.
The rTree is an infrastructure for prefix-matching features in LLM multi-turn dialogue scenarios. 

We will present the usage of these data structures with the vTensor operations in the next subsections.

\subsection{vTensor Operation}
The vTensor Operation provides the memory-related operations for vTensor, including three main modules: allocation, deallocation, and advanced operation for manipulating the prefix tree.
\Tbl{tab:operations} shows vTensor operations.


\begin{table}[t]
\centering
    \begin{tabular}{c|c}
    \hline
    \textbf{Modules} & \textbf{vTensor Operations} \\ \hline
    \textbf{Allocation} & \textit{vAlloc, pAlloc, Map} \\ \hline
    \textbf{Deallocation} & \textit{vFree, pFree, Unmap} \\ \hline
    \textbf{Tree Operation} & \textit{rPush, rPrefixMatch} \\ \hline
    \end{tabular}
    \caption{The categories of vTensor operations.}
    \label{tab:operations}
    \vspace*{-5mm}
\end{table}

\subsubsection{Allocation} 

\paragraph{GPU Interface}
As shown in \Fig{fig:overview}, the GPU virtual memory management (VMM) API is the fundamental interface between the vTensor and GPU. 
The VMM allocation occurs with three key primitive operations. 
First, address reservation (\texttt{cuMemReserve}): This initial step involves determining the required memory size and securing the appropriate virtual address space. 
Next, physical creation (\texttt{cuMemCreate}): Here, the system generates actual data segments in GPU physical memory. 
Finally, mapping (\texttt{cuMemMap}): This stage links the physical data segments to the reserved virtual address, allowing tensors to access the memory efficiently. This trio of steps, reserving, creating, and mapping, forms the comprehensive mechanism for allocating a primitive virtual memory block.
VTO invokes these primitive APIs through operations to support diverse KV cache management.

\paragraph{pAlloc($N$)}
The physical allocation {pAlloc($N$)} first will check the pSet to find available physical chunks without storing the tokens.
Suppose the rest of the available chunks in pSet are $M$, which is less than $N$. In that case, the {pAlloc} will continue to utilize the \texttt{cuMemCreate} API to allocate new $N - M$ physical chunks from the GPU global memory and insert returned handles into pSet. 
In this study, we use 2~MB chunk size.
It's the only vTensor operation that can increase GPU memory usage.
Physical allocation is the basic memory operation that is independent of virtual allocation.

\paragraph{vAlloc($S$)}
The {vAlloc($S$)} is the virtual allocation function, which similarly checks the vSet to find an available virtual address space, else allocate and reserve the virtual address with $S$ bytes of virtual address space via \texttt{cuMemAddressReserve} API. 
In an LLM task, we allocate the virtual address with the $S =$ maximum sequence length, and all virtual addresses share the same length.
It's a lightweight operation that only reserves the virtual address space for the vTensor and does not allocate any physical chunks.
The virtual address will be recorded in the vSet.   
In contrast, the virtual allocation does not allocate new physical memory. 
Under the massive overhead~\cite{guo2024gmlake} caused by CUDA VMM APIs, the decoupling virtual and physical allocation design is exactly the key design for efficient KV cache defragmentation.

\paragraph{Map(VA, PCs)}
After allocating the virtual address (VA) and physical chunks (PCs), we need to map the virtual address to a subset of physical chunks and make the virtual address accessible to the real data using the \texttt{cuMemMap} API.
The mapping function is the key to ensuring our design is flexible to memory management.

\begin{figure*}[t]
    \begin{center}
    \includegraphics[width=1\textwidth]{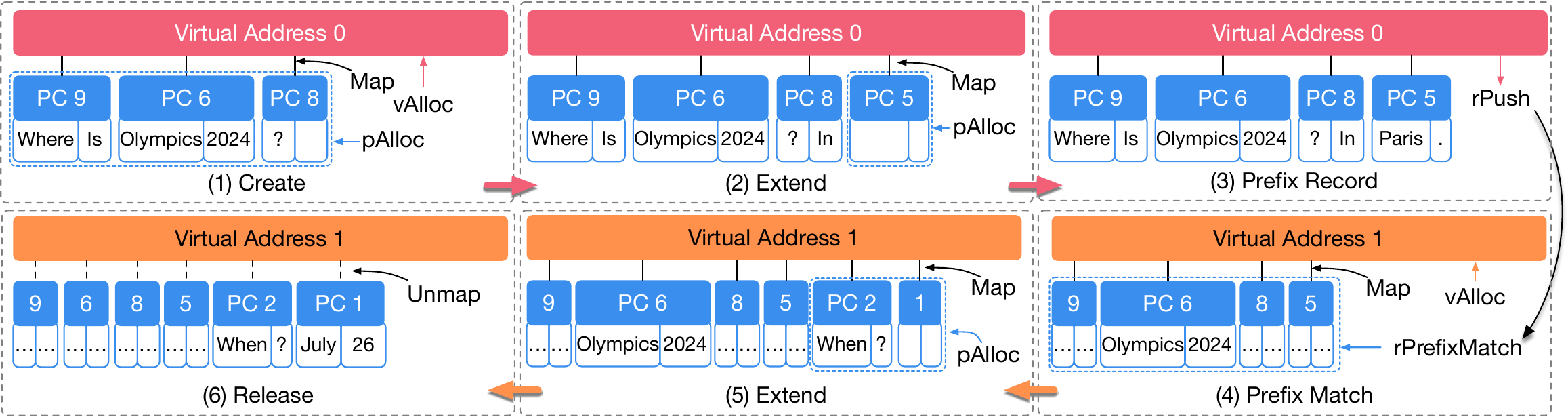}
    \caption{An example of vTensor scheduling.}
    \label{fig:vts}
    \end{center}
    \vspace{-5mm}
\end{figure*}

\subsubsection{Deallocation.} 
The operation in deallocating is the reverse operation corresponding to the allocation.
The state of the unmapped physical chunks and virtual addresses will be modified to make them available for future reuse.
The \textbf{Unmap()} operation is called to unmap the virtual address from physical chunks, and \textbf{vFree()} and \textbf{pFree()} are to free the resources on GPU. 
 The deallocation module introduces a \textbf{lazy deallocation} mechanism instead of actively deallocating the vTensor and physical chunks to reduce memory operation overheads.
In practice, we usually use the unmap operation instead of freeing their resource on the GPU until the serving task ends.



\subsubsection{Tree Operations}
The tree operations are to manipulate the prefix tree. 
The \textbf{rPush(vTensor)} operation is used to insert the vTensor pointer into the rTree for prefix matching features.
In our rTree design, we store each vTensor as a distinct node. 
First, rTree will check if there are any nodes that share the common prefix pattern with the inserting vTensor. 
If there is, we will keep traversing its children nodes until none match the remaining prefix, at which point the vTensor will be inserted as a child of the current node.
If there are no such nodes, then this vTensor will be inserted as the root of a new radix tree in the rTree. 

The \textbf{rPrefixMatch(vTensor)} operation is used to search the rTtree for a node that matches the prefix pattern of the vTensor.
The rTree traversing process is similar to the rPush operation. 
The rPrefixMatch will return the vTensor pointer stored and still maintain this prefix record for subsequent requests.

\subsection{vTensor Scheduler}
\label{sec:vTS}

This section presents the vTensor Scheduler (VTS), which leverages vTensor Pool and vTensor Operation to achieve efficient memory scheduling.
The vTensor Scheduler is designed to receive memory instructions from the \proj{} scheduler and execute the vTensor operations. \Fig{fig:vts} shows the VTS scheduling strategy with an example.
The vTensor scheduler has five basic operations, including create, extend, related, and two prefix-related operations: prefix record and prefix match.

\paragraph{Create}
The Create is to initialize the new request.
It will employ the pAlloc and vAlloc to allocate needed physical chunks and virtual memory addresses.
Then, VTS will map the physical chunks of the virtual memory using the Map operation, combining a complete vTensor, which will return to the CUDA kernel.

\paragraph{Extend}
As shown in \Fig{fig:vts} (2), the Extend operation is fundamental to LLM serving of FlexInfer. 
VTS dynamically expands sequence length during decoding to generate new tokens iteratively.
VTS uses the pAlloc to allocate a new physical chunk and map it to the virtual address, which has been allocated in the Create operation. 
This dynamic extension policy is crucial to vTensor's design, enabling memory flexibility and computational efficiency. 
The process involves two key advantages. 

First, an extension of a new physical chunk in the KV cache is dispatched preemptively before computation with existing chunks. 
This asynchronous approach allows memory allocation to overlap with computation, minimizing overhead. 
Notably, pAlloc may reuse available physical chunks from the pSet, potentially avoiding new GPU allocations. 
These scheduling and optimization techniques significantly reduce memory allocation overhead compared to previous methods. 
Second, vTensor consistently maintains compact GPU memory usage, contrasting with prior KV cache systems that pre-allocated all GPU memory exclusively for the KV cache.

\paragraph{Prefix Record and Match}
Prefix Record and Match operations support the prefix KV cache, reducing recomputation in multi-turn dialogues. As illustrated in \Fig{fig:vts}(3), when a dialogue is marked for continuation, VTS records the vTensor in the rTree as a prefix candidate. 
\Fig{fig:vts}(4) demonstrates how VTS efficiently retrieves the vTensor using a radix tree when a subsequent request requires the prefix.
Upon obtaining the prefix vTensor (VA-0), VTS allocates a new virtual address (VA-1) and duplicates the page table and metadata of VA-0. Importantly, physical chunks are not reallocated; instead, they are mapped to existing chunks at the new virtual address (VA-1). 
This approach significantly reduces memory allocation overhead and usage, enabling efficient support for the prefix KV cache.
Subsequently, the prefix-sharing request transitions to a regular request, utilizing the Extend mechanism for generating additional tokens, as depicted in \Fig{fig:vts}(5).

\paragraph{Release}
Fundamentally, our VTM design retains GPU resources until either the LLM serving ends or an explicit memory-emptying operation is invoked, which can occur at any time. 
This approach allows VTM to maintain allocated resources, minimizing memory overhead. 
However, a Release policy is still necessary to return virtual addresses and physical chunks to their available status, enabling other requests to use them, as shown in \Fig{fig:vts}(6).



%% file: tex/felxinfer.tex
\section{FlexInfer Scheduler}
\label{sec:flex-sche}
FlexInfer enables efficient LLM execution on GPU tensor cores by scheduling highly optimized kernels without memory addressing constraints. 
The FlexInfer scheduler translates diverse request types into distinct memory instructions, which are then processed by the VTM. These instructions are detailed in \Sec{sec:vTS}.
Algorithm~\ref{alg:request-scheduling} summarizes FlexInfer's request scheduling strategy, which asynchronously dispatches computation and memory actions. 
For new requests, FlexInfer invokes Create or Prefix Match operations in the VTM to initialize request memory and construct the vTensor. 
Subsequently, it provides the vTensor pointer to the computation kernel and asynchronously dispatches it to the GPU tensor core.
During decoding, it dynamically extends the vTensor's physical chunks to meet memory demands. 
If resources are insufficient, the scheduler preempts the lowest priority request and reallocates its resources.

The scheduler employs a series of vTensor scheduling strategies to interact with the VTM, streamlining the scheduling logic. Upon program termination, the scheduler releases all vTensors and associated resources.
This asynchronous scheduling ensures that extension operations precede computation kernels, effectively overlapping and concealing memory overhead during LLM serving, thus enhancing computation and memory flexibility.

 \begin{algorithm2e}[t]
        \small
           \DontPrintSemicolon
           \KwIn{
               configurations for LLM serving: $config$;\\
               \ \ \ \ \ \ \ \ \ \ \ \ \ prioritized LLM serving request queue \\
               \ \ \ \ \ \ \ \ \ \ \ \ \ requests: ${Requests}$. \\
               \ \ \ \ \ \ \ \ \ \ \ \ \ vTensor Manager: $VTM$.
           }
           \KwOut{
               Requests status, Scheduled Requests.
           }
       
           \SetKwFunction{FMain}{Prefill} 
           \SetKwProg{Fn}{def}{:}{}
           \Fn{\FMain{request, config}}{
            \If{prefix} 
            {   
                status = VTM.call(\textbf{Prefix\_Match}, Request) \\
                \If{status == failure} {
                    VTM.call(\textbf{Create}, Request) \\
                    VTM.call(\textbf{Extend}, Request) \\
                    VTM.call(\textbf{Extend}, Request)  \tcp*[h]{Pre Extend for the next physical chunk}
                }
            }\Else{
                VTM.call(\textbf{Create}, Request) \\
                VTM.call(\textbf{Extend}, Request) \\
                VTM.call(\textbf{Extend}, Request)
            }
           }    
           \SetKwFunction{FMain}{Decode} 
           \SetKwProg{Fn}{def}{:}{}
           \Fn{\FMain{request, config}}{
                    \If{no enough resources} {
                        Preempted\_Request = Get\_Preempted\_Request() \\
                        VTM.call(\textbf{Release}, Preempted\_Request)
                    }
                    VTM.call(\textbf{Extend}, Request) \tcp*[h]{Extend for the next physical chunks}
           }

           \SetKwFunction{FMain}{Schedule} 
           \SetKwProg{Fn}{def}{:}{}
           \Fn{\FMain{config, Requests}}{
                \ForEach{$request \in Requests$}{ 
                    \If{Request.type == Prompt} { 
                        Prefill(request, config)
                    }
                    \ElseIf{Request.type == Decode} 
                    {
                        Decode(request, config)
                    }
                    \ElseIf{Request.type == Release} {
                        VTM.call(\textbf{Release}, Request)
                    }
                    \ElseIf{Request.type == Finished} {
                        VTM.call(\textbf{Prefix Record}, Request)
                    }
    
                    \Return Record\_Request\_Status(Requests)
                }
           }
           \caption{The \texttt{request scheduling} Strategy.}
           \label{alg:request-scheduling}
    \end{algorithm2e}

%% file: tex/evaluation.tex
\section{Evaluation}
We develop \proj{} based on the vLLM project with 3000 lines of Python and C++ modifications, together with a C++ library for vTensor manager, and integrate flash-attn-2.5.8 into \proj{} using vTensor.
We evaluate the performance of \proj{} from computation kernels to end-to-end serving scenarios with popular language models to verify computation and memory flexibility. We choose vLLM v0.4.2 as the baseline for comparison under diverse conditions,
including computation kernels in both decoding and prefix-prefilling stages, multi-turn chatbot, single-generation, and prefix-sharing scenarios. Through the evaluation, we aim to demonstrate the effectiveness of \proj{} in improving the flexibility of LLM serving systems.
\proj{} achieves an average performance improvement of 2.12$\times$ and 3.15$\times$ in the decoding kernel and prefix-prefilling kernels respectively, which is up to 3.27 $\times$ and 3.92$\times$. End-to-end evaluation illustrates 1.86$\times$ average throughput improvement and 
up to $2.4\times$ throughput increase in end-to-end serving scenarios compared to vLLM. Moreover, it saves up to 80\% (57~GB) GPU memory usage compared to vLLM. 
    
\subsection[short]{Methodology}    
\paragraph[short]{Setup.} We conduct all experiments on a server with eight NVIDIA A100(80GB) GPUs and an Intel Xeon Platinum 8369B CPU. These GPUs are interconnected via NVLink. 

\paragraph[short]{Evaluation Scenarios.} We evaluate \proj{} across multiple self-attention computation kernels and end-to-end serving scenarios on different LLMs. For end-to-end serving, we focus on single-generation and prefix-caching scenarios containing multi-turn chatbot and prefix-sharing scenarios.
For kernel implementations, we focused on a self-attention mechanism, including decoding and prefix-prefilling kernels. This mechanism is evaluated in GQA with the KV head number equal to 4 and Q head=32 as default to align with the LLMs we evaluate.

\paragraph[short]{Models and Dataset.} We conduct our evaluation on three representative LLMs with their official configurations, including Yi-6B-200K (32 heads for query, 4 heads for KV), Yi-9B-32K (32 heads for query, 4 heads for KV), and Yi-34B-32K (56 heads for query, 8 heads for KV)~\cite{ai2024yi,Yi-1.5-34B-32K}.
We have also evaluated LLama models~\cite{touvron2023llama,touvron2023llama2,llama3modelcard}, which have the same model architecture as the Yi-series model.
We only report the results in Yi models owing to the space limitation. 
We use a synthetic dataset from SGLang~\cite{zheng2024sglang}
for multi-turn chatbot and prefix-sharing scenarios. In the context of single-generation scenarios, we construct our dataset by excerpting data from an open-source LLM long context dataset named LV-Eval~\cite{yuan2024lveval}. To emulate real-world complexity and challenge models with long input-output, we truncate the dataset such that
prompts range in length from 6,000 to 8,000 tokens, while output lengths fluctuate between 2,000 and 4,000 tokens, thereby avoiding a uniform alignment between prompt and response sizes.    
For multi-turn chat, we evaluate each request with distinct prompts, which means less common prefix-sharing and focuses on multi-turn chatbot performance. 
The prompt length is set to 2K and generates 2K tokens for each turn. For the prefix-sharing scenario, each request shares the same prefix prompt with 12K length and 4k distinct prompt and generates 10 tokens. 

\paragraph[short]{Baselines.}  
In end-to-end evaluation, we compare \proj{} against vLLM v0.4.2~\cite{vLLM042} with both single-turn chat and prefix-caching scenarios. For kernel evaluation, we compare \proj{} against different versions of FlashAttention, PagedAttention from vLLM, and triton prefix-prefilling kernel from SGLang~\cite{zheng2024sglang}, which is adopted by vLLM, too.
For convenience, we take PagedAttention as vLLM, FlashAttention with paged KV cache as FA\_paged, our virtual-memory-enabled FlashAttention as \proj, native FlashAttention as FA\_native and triton prefix-prefilling kernels as SGLang. We adopt the v2.5.8 version of FlashAttention.
   
\subsection[short]{Computation Kernel Performance Evaluation}

\begin{figure}[htbp]
    \centering
    \includegraphics[width=\linewidth]{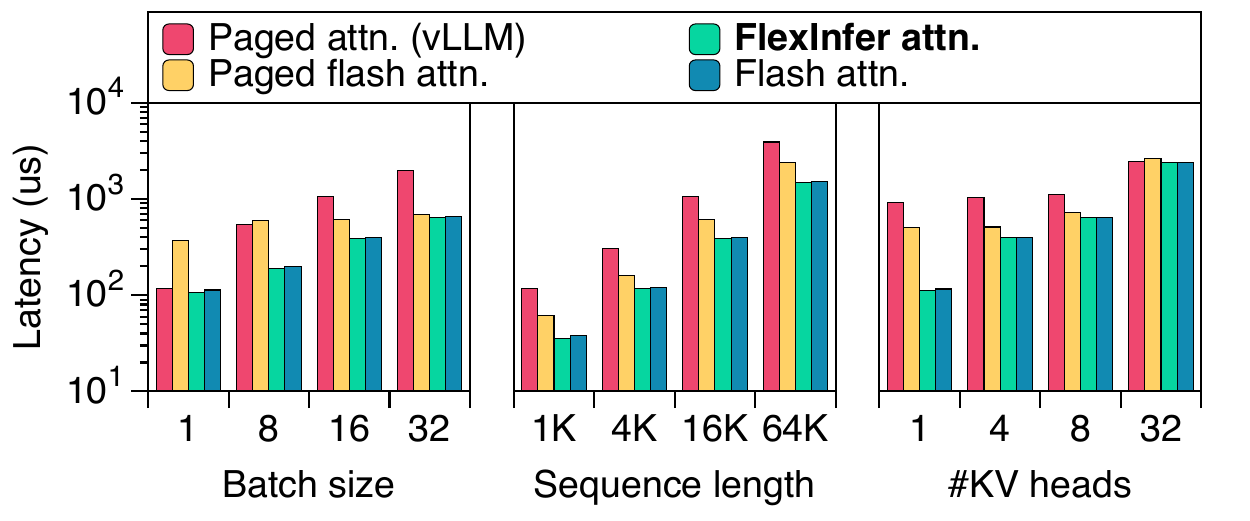}
    \caption{Evaluation on decoding kernels from different perspectives. The left and middle figure illustrates the effect on decoding kernels when batch size and KV cache sequence length vary. The right one 
    shows how the attention head numbers of K and V in the self-attention mechanism affect the performance of decoding kernels.}
    \label{fig:eval-kernels}
    \vspace*{-3mm}
\end{figure}

\subsubsection{Decoding Kernel Evaluation}

In the left of \Fig{fig:eval-kernels}, we evaluate different attention kernels with varied batch sizes with KV cache sequence length fixed to 16K. 
We observe that \proj{} owns an average speedup of 2.78$\times$ compared to vLLM PagedAttention and 1.02$\times$ performance compared to native FlashAttention. 
This acceleration reaches the peak when batch size comes to 32, which is 3.08$\times$ speedup compared to PagedAttention and 1.2$\times$ compared to native FlashAttention.
We can also notice that the performance of Paged FlashAttention stays the same in the evaluated batch size range, and the average overhead of Page FlashAttention is 42\% compared to \proj{}. 
We assume that it's caused by supporting paged KV cache memory accessing, as the native FlashAttention performs normally.
This phenomenon to some extent indicates the difficulty of refactoring the paged kernels to support the architecture with higher compute capability while showcasing the superior performance and computation flexibility of \proj{}

For the middle of \Fig{fig:eval-kernels}, we conduct kernel evaluation on different sequence lengths of the KV cache with batch size = 16. 
When the sequence length varies, we can observe an average performance improvement in \proj{} kernels of 2.67$\times$, 1.59$\times$ and 1.02$\times$ speedup in comparison with PageAttention, Paged FlashAttention and native FlashAttention respectively.
The peak speedup occurs when the sequence length is equal to 1024, which is 3.27$\times$ higher than PageAttention.

In the right of the \Fig{fig:eval-kernels}, we exploit the performance of decoding kernels with different KV head numbers. For convenience, the batch size and KV cache sequence length are set to 16 and 16K respectively.
We can notice a trend that the speedup of \proj{} kernels increases as the KV head numbers. To be specific, this speedup increases from 1.03$\times$ in MHA when the KV head number is equal to 32 to 8.0$\times$ in MQA when the KV head number is equal to 1. 
Within this process, \proj{} kernel achieves 1.01$\times$ of native FlashAttention performance and 1.23$\times$ of Paged FlashAttention performance, proving its robustness and computation flexibility.

\begin{figure}[t]
    \centering
    \includegraphics[width=\linewidth]{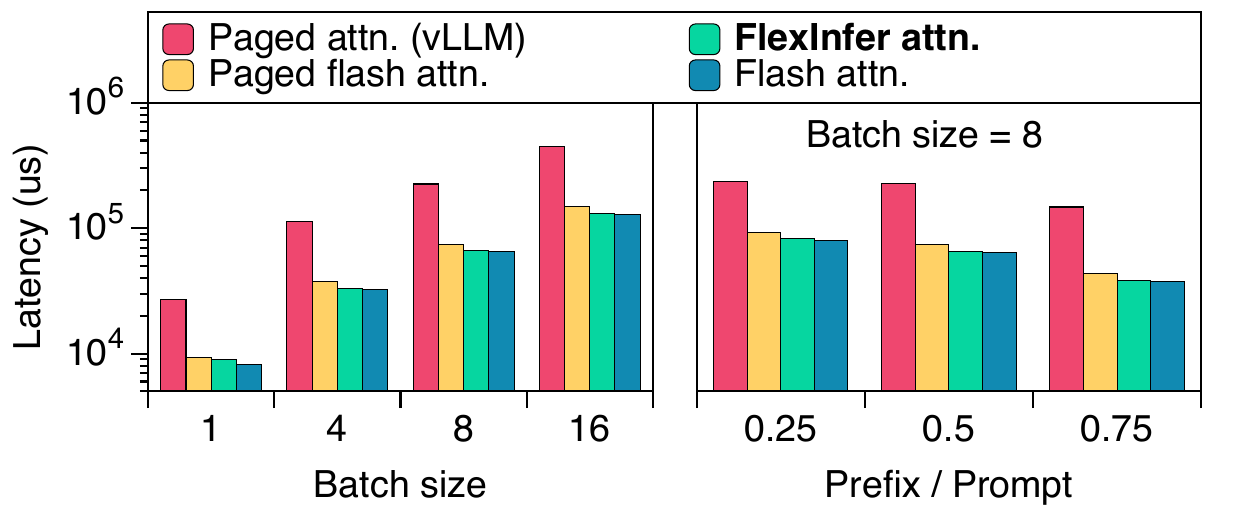}
    \caption{Evaluation on prefix-prefilling kernel implementations. The left figure explores the effect on prefix-prefilling kernels with varied batch sizes. 
    The right figure shows how prefix-prefilling kernels perform with different prefix ratios.}
    \label{fig:eval-prefilling}
    \vspace*{-5mm}
\end{figure}

\subsubsection{Prefix-prefilling Kernel Evaluation}

On the left of \Fig{fig:eval-prefilling}, we evaluate different prefix-prefilling kernels with varied batch sizes. The sequence length is fixed to 16K and the prefix ratio is 0.5.
\proj{}'s prefix-prefilling kernel achieves $3.40\times$ speed up than the triton kernel proposed by SGLang and adopted by vLLM while preserving 0.98 $\times$ performance of native FlashAttention and 1.12$times$ of Paged FlashAttention.
The speed-up grows with batch size, because of the higher compute capability of \proj{} kernels, and reaches peak when batch size is 16, which is 3.49$\times$ compared to triton kernel. 

In the right of \Fig{fig:eval-prefilling}, with the decrease of the prefix: prompt ratio, we witness the rise in speed-up from 2.9$\times$ to 3.92$\times$. In the case of a fixed context length, as the proportion of prefix tokens increases, the computational power required for attention operations will be relatively small. At this point, \proj{} and flash-attn can fully utilize the computational power of the GPU. 

\subsubsection{Page Table Overhead Analysis}
We can observe the obvious page table overhead from the above analysis on \Fig{fig:eval-kernels} and \Fig{fig:eval-prefilling}. In our evaluation, the performance of FA\_paged is fluctuant and can lead to up to 3x more latency than the native one. And we attribute this phenomenon to FA\_paged's bad compatibility with this input shape.
When the input scale gets larger, this overhead stabilizes to an average of 43\% in the decoding kernel and 15\% in the prefix-prefilling kernel. PageAttention, however, optimized specially for paged KV cache memory accessing, suffers the difficulty of 
refactoring the kernels to support the architecture with higher compute capability instead of this fluctuation in performance. Native FlashAttention, on the other hand, leveraging advanced computing capability and achieving up to 3.2x speed up compared to PageAttention, takes evident overhead to migrate its implementation to support paged KV cache. 
With \proj{}'s computation flexibility, we can easily integrate with efficient CUDA kernels, significantly improving the performance based on GPU tensor core with little computation overhead.


\subsection[short]{End-to-End Performance Evaluation}
\label{sec:computation-e2e-evaluation} 
    \begin{figure}[t]
        \centering
        \includegraphics[width=\linewidth]{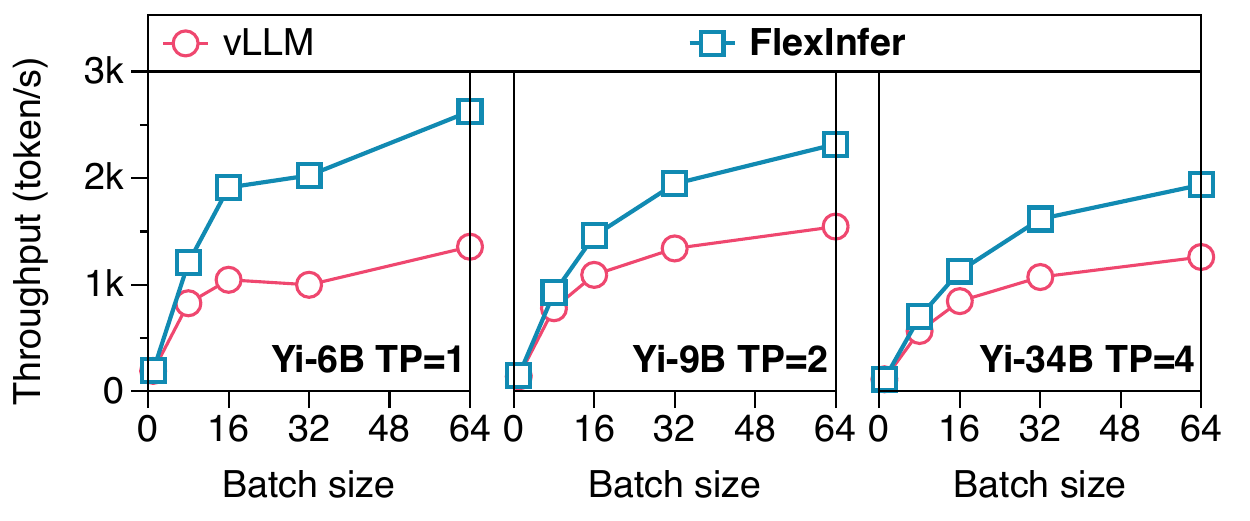} 
    \vspace{-5mm}
        \caption{End-to-end Single-generation Evaluation on Computation Flexibility with different models and parallelisms.}
        \label{fig:eval-compflex-e2e-1}
        \vspace{-3mm}
    \end{figure}

    \subsubsection{Single-generation Scenario.} In single-generation, one request demands one prefilling computation and multiple decoding computations. Therefore, we conduct evaluations to assess the impact of efficient decoding kernels in end-to-end scenarios.
    We evaluate the single-generation scenario with Yi-6B-200K, Yi-9B-32K, and Yi-34B-32K, with tensor parallelism degrees of 1, 2, and 4 respectively.

    We observe a substantial throughput increase for \proj{} in our evaluation. As illustrated in \Fig{fig:eval-compflex-e2e-1}, the throughput of \proj{} exhibits a notable upward trend which is on average 1.8 $\times$, 1.3$\times$ and 1.4$\times$ respectively for 3 models as the batch size grows.
    These models achieve peak throughout when batch size is 64, which is 2.02$\times$, 1.5$\times$, and 1.53$\times$ performance compared to vLLM. 
    This enhancement is directly attributed to the joint effort of optimized decoding kernels and memory optimizations within \proj{}.
    
    \subsubsection{Prefix-caching Scenarios.} Prefix-caching scenarios include multi-turn chat and prefix-sharing scenarios. 
    For multi-turn chat evaluation, we focus on the effectiveness of end-to-end performance of \proj{}. For fork evaluation, we focus on the effectiveness of prefix-prefilling kernel. 

\begin{figure}[t]
        \centering
        \includegraphics[width=\linewidth]{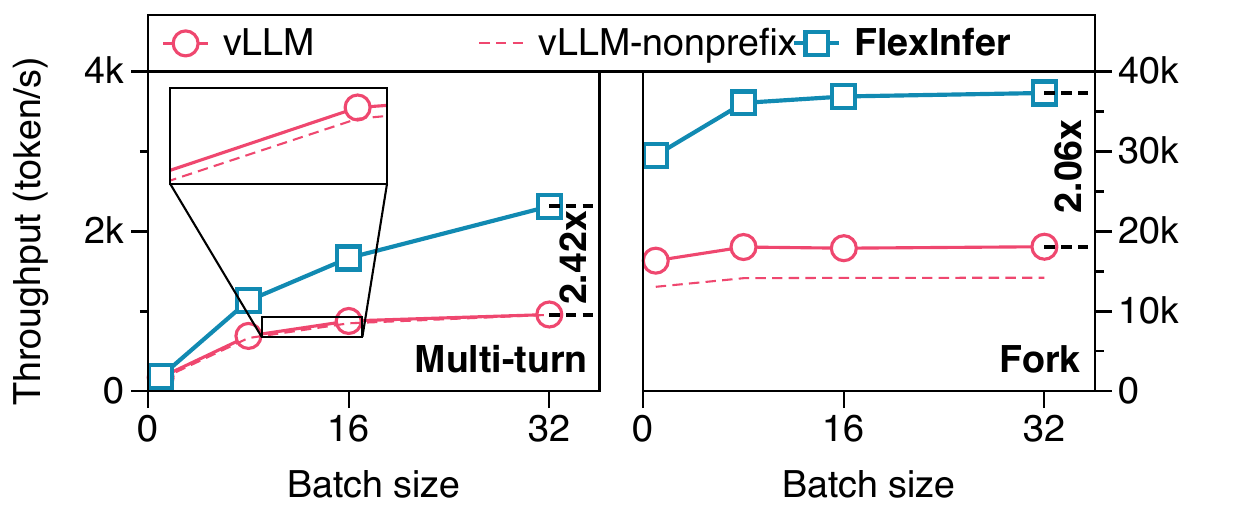} 
        
    \vspace{-3mm}
        \caption{End-to-end Prefix-caching Evaluation on Computation Flexibility with Yi-6B.}
        \label{fig:eval-compflex-prefix-1}
        \vspace{-5mm}
    \end{figure}

    As the right of \Fig{fig:eval-compflex-prefix-1} shows, prefix cache achieves acceleration for both \proj{} and vLLM when computation demands are low. However, \proj{} outperforms vLLM by up to 2.0x in throughput when batch size grows. 
    \proj{} achieves this remarkable acceleration primarily through its highly efficient prefix-prefilling kernel. As the left of \Fig{fig:eval-compflex-prefix-1} shows, due to the poor implementation of the prefix-prefilling kernel used in vLLM, its performance is consistent with not using the prefix kernel and directly using the flash-attn kernel. But, \proj{} can outperform vLLM by up to 2.42x in throughput.


\subsection[short]{Memory Efficiency of FlexInfer}
\label{sec:memory-flexibility-evaluation}


    We evaluate memory flexibility by collecting \proj{} runtime memory trace when serving requests with Yi-6B in A100 (80G).
    \Fig{fig:eval-memflex-1} depicts the relationship between running batch size and memory usage of \proj{} and vLLM. \proj{} can adjust the memory usage according to the batch size. When the batch size is small, \proj{} can save almost all KV cache GPU memory compared to prior works.

    \Fig{fig:eval-memflex-3} takes a more practical perspective. The existing solution statically partitions the GPU memory for KV cache memory management.
    However, when the request rate fluctuates, the memory usage of the KV cache is not fully utilized, leading to a waste of GPU memory.
    \proj{} provides flexibility in memory reservation and enables future application colocation during LLM serving.



%% file: tex/relatedwork.tex
\section{Related Works and Discussion}\label{sec:relatedworks}

\paragraph*{Virtual Memory Management.} 
{Virutal memory management has been long considered in machine learning systems to boost memory utilization and realize unified memory management.
NVIDIA's Unified Virtual Memory (UVM)~\cite{uvm6} technology enables applications to allocate memory accessible for both reading and writing by code executing on CPUs or GPUs, a feature supported beginning with the Pascal GPU architecture~\cite{pascalgpu}. 
vDNN~\cite{vDNN} propose a runtime memory management solution that virtualizes the memory usage of deep neural networks across both GPU and CPU memories to reduce memory usage in DNN workloads. 
In the LLM era, GMLake~\cite{guo2024gmlake} employs CUDA low-level virtual memory management to manage the memory of large-scale models finetuning workloads and reduces memory fragmentation.
For serving scenarios, vAttention~\cite{prabhu2024vattention} is a concurrent work that focuses on dynamically managing KV cache with similar CUDA low-level virtual memory management APIs and facilitates efficient LLM serving.
}
\paragraph*{Efficient LLM serving techniques.}

 \begin{figure}[t]
    \centering
    \subfloat[\textbf{Memory Consumption}.]{
    \label{fig:eval-memflex-1}
    \includegraphics[width=0.46\linewidth]{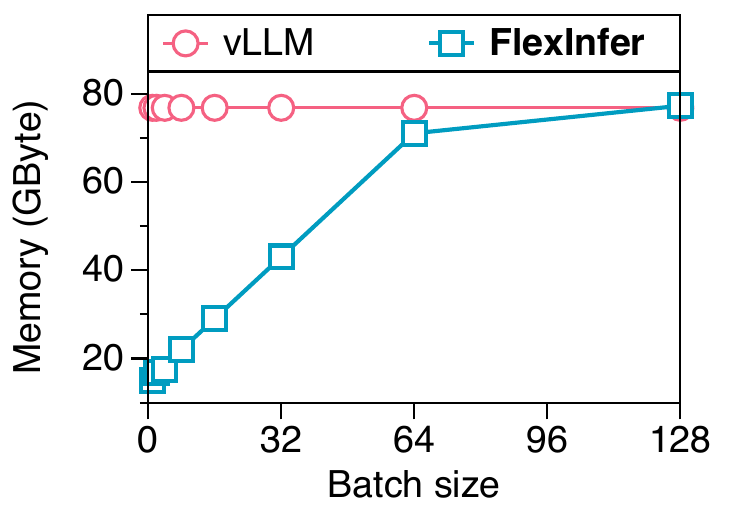}}
    \hfill
    \subfloat[\textbf{Memory Trace.}]{
    \label{fig:eval-memflex-3}
    \includegraphics[width=0.46\linewidth]{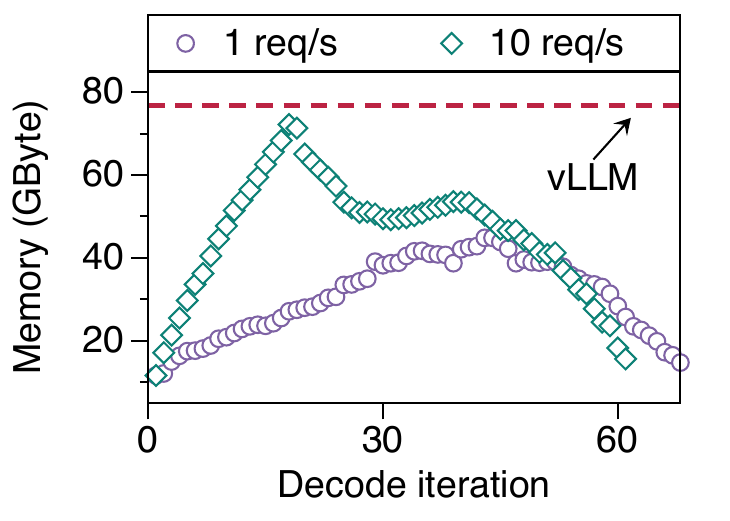}} 
    \hfill
    \vspace*{\fill} 
    \caption{Memory flexibility evaluation for Yi-6B model. (b) shows memory traces for different request rates.}
\vspace{-5mm}
    \label{fig:eval-memflex}
\end{figure}

{Memory efficient techniques like Quantization~\cite{zhao2024atom, guo2022squant, AWQ, Olive,han2015deep, guo2022ant, xiao2023smoothquant} and Pruning~\cite{guan2024fractal,guo2020accelerating,wang2021dual,liu2023dejavucontextualsparsity, guo2024accelerating} serve as important roles to reduce the memory footprint of LLM workloads.
Explorations in computation such as FlashAttention~\cite{FlashAttention, FlashAttention2}, FlashInfer~\cite{FlashInfer} improve the computation efficiency of LLM serving.
There are also discussions on LLM serving scheduling strategies. Continuous batching~\cite{280922, vLLM} is proposed to implement dynamic batch partitioning and mitigate the interference between different requests.
To further eliminate the interference between the prefill and decode stage, recent works~\cite{zhong2024distserve,gao2024costefficientlargelanguagemodel,patel2024splitwise,hu2024tetri} propose to disaggregate the prefill and decode stage into different instances.  
Efficient KV cache management has been also a hot topic recently. KV cache defragmentation~\cite{vLLM} facilitates larger serving batch sizes in the serving system. 
Prefix-caching~\cite{gim2024promptcache,zheng2024sglang} enables faster prefill processing by managing and reusing the previous KV cache.
Distributed KV cache~\cite{lin2024infinitellm} supports efficient KV cache orchestration across the memory of GPUs and CPUs in a data center.
Different self-attention types such as GQA~\cite{ainslie2023gqa}, MQA~\cite{shazeer2019fast}, and MLA~\cite{deepseekai2024deepseekv2} are designed to further reduce the volume of KV cache, which is orthogonal to general compression techniques.
}

%% file: tex/conclusion.tex
\section{Conclusion}\label{sec:conclusion}
By introducing the vTensor abstraction and decoupling memory management from computation, FlexInfer has addressed critical inefficiencies in existing LLM serving systems. The \proj{}'s computation flexibility facilitates up to 3.92$\times$ kernel performance improvements compared to widely adopted kernel implementation in the LLM serving system.
Our framework achieves up to 2.4$\times$ and on average 1.86 $\times$  speedup across diverse models and applications in end-to-end experiments. Moreover, FlexInfer's ability to free 71.25\% (57GB) of GPU memory on average opens new possibilities for memory-intensive tasks. 
These improvements in both computational efficiency and memory utilization mark a crucial step forward for LLM deployment. 
As LLMs grow in size and importance, our framework would be essential to ensure their efficient and cost-effective operation at scale.

%% file: main.bbl
\begin{thebibliography}{10}

\bibitem{Yi-1.5-34B-32K}
01-ai.
\newblock Yi-1.5-34b-32k model card, 2024.

\bibitem{agrawal2024taming}
Amey Agrawal, Nitin Kedia, Ashish Panwar, Jayashree Mohan, Nipun Kwatra, Bhargav~S. Gulavani, Alexey Tumanov, and Ramachandran Ramjee.
\newblock Taming throughput-latency tradeoff in llm inference with sarathi-serve, 2024.

\bibitem{ai2024yi}
01. AI, Alex Young, Bei Chen, Chao Li, Chengen Huang, Ge~Zhang, Guanwei Zhang, Heng Li, Jiangcheng Zhu, Jianqun Chen, Jing Chang, Kaidong Yu, Peng Liu, Qiang Liu, Shawn Yue, Senbin Yang, Shiming Yang, Tao Yu, Wen Xie, Wenhao Huang, Xiaohui Hu, Xiaoyi Ren, Xinyao Niu, Pengcheng Nie, Yuchi Xu, Yudong Liu, Yue Wang, Yuxuan Cai, Zhenyu Gu, Zhiyuan Liu, and Zonghong Dai.
\newblock Yi: Open foundation models by 01.ai, 2024.

\bibitem{llama3modelcard}
AI@Meta.
\newblock Llama 3 model card.
\newblock 2024.

\bibitem{ainslie2023gqa}
Joshua Ainslie, James Lee-Thorp, Michiel de~Jong, Yury Zemlyanskiy, Federico Lebrón, and Sumit Sanghai.
\newblock Gqa: Training generalized multi-query transformer models from multi-head checkpoints, 2023.

\bibitem{bai2022training}
Yuntao Bai, Andy Jones, Kamal Ndousse, Amanda Askell, Anna Chen, Nova DasSarma, Dawn Drain, Stanislav Fort, Deep Ganguli, Tom Henighan, Nicholas Joseph, Saurav Kadavath, Jackson Kernion, Tom Conerly, Sheer El-Showk, Nelson Elhage, Zac Hatfield-Dodds, Danny Hernandez, Tristan Hume, Scott Johnston, Shauna Kravec, Liane Lovitt, Neel Nanda, Catherine Olsson, Dario Amodei, Tom Brown, Jack Clark, Sam McCandlish, Chris Olah, Ben Mann, and Jared Kaplan.
\newblock Training a helpful and harmless assistant with reinforcement learning from human feedback, 2022.

\bibitem{brown2020language}
Tom Brown, Benjamin Mann, Nick Ryder, Melanie Subbiah, Jared~D Kaplan, Prafulla Dhariwal, Arvind Neelakantan, Pranav Shyam, Girish Sastry, Amanda Askell, et~al.
\newblock Language models are few-shot learners.
\newblock {\em Advances in neural information processing systems}, 33:1877--1901, 2020.

\bibitem{GPT3}
Tom~B. Brown, Benjamin Mann, Nick Ryder, et~al.
\newblock Language models are few-shot learners.
\newblock In {\em Advances in Neural Information Processing Systems 33: Annual Conference on Neural Information Processing Systems 2020, NeurIPS 2020, December 6-12, 2020, virtual}, 2020.

\bibitem{vicuna2023}
Wei-Lin Chiang, Zhuohan Li, Zi~Lin, Ying Sheng, Zhanghao Wu, Hao Zhang, Lianmin Zheng, Siyuan Zhuang, Yonghao Zhuang, Joseph~E. Gonzalez, Ion Stoica, and Eric~P. Xing.
\newblock Vicuna: An open-source chatbot impressing gpt-4 with 90\%* chatgpt quality, March 2023.

\bibitem{cuBLAS}
NVIDIA Corporation.
\newblock cublas library.
\newblock \url{https://developer.nvidia.com/cublas}.

\bibitem{nvidia2024cuda}
NVIDIA Corporation.
\newblock Cuda driver api documentation.
\newblock \url{https://docs.nvidia.com/cuda/cuda-driver-api/group__CUDA__VA.html#group__CUDA__VA}.

\bibitem{nvidia2024vm}
NVIDIA Corporation.
\newblock Introducing low-level gpu virtual memory management.
\newblock \url{https://developer.nvidia.com/blog/introducing-low-level-gpu-virtual-memory-management/}.

\bibitem{cui2023survey}
Can Cui, Yunsheng Ma, Xu~Cao, Wenqian Ye, Yang Zhou, Kaizhao Liang, Jintai Chen, Juanwu Lu, Zichong Yang, Kuei-Da Liao, Tianren Gao, Erlong Li, Kun Tang, Zhipeng Cao, Tong Zhou, Ao~Liu, Xinrui Yan, Shuqi Mei, Jianguo Cao, Ziran Wang, and Chao Zheng.
\newblock A survey on multimodal large language models for autonomous driving, 2023.

\bibitem{cui2022m6rec}
Zeyu Cui, Jianxin Ma, Chang Zhou, Jingren Zhou, and Hongxia Yang.
\newblock M6-rec: Generative pretrained language models are open-ended recommender systems, 2022.

\bibitem{FlashAttention2}
Tri Dao.
\newblock Flashattention-2: Faster attention with better parallelism and work partitioning.
\newblock {\em CoRR}, abs/2307.08691, 2023.

\bibitem{FlashAttention}
Tri Dao, Daniel~Y. Fu, Stefano Ermon, Atri Rudra, and Christopher R{\'{e}}.
\newblock Flashattention: Fast and memory-efficient exact attention with io-awareness.
\newblock In {\em NeurIPS}, 2022.

\bibitem{deepseekai2024deepseekv2}
DeepSeek-AI, Aixin Liu, Bei Feng, Bin Wang, Bingxuan Wang, Bo~Liu, Chenggang Zhao, Chengqi Dengr, Chong Ruan, Damai Dai, Daya Guo, Dejian Yang, Deli Chen, Dongjie Ji, Erhang Li, Fangyun Lin, Fuli Luo, Guangbo Hao, Guanting Chen, Guowei Li, H.~Zhang, Hanwei Xu, Hao Yang, Haowei Zhang, Honghui Ding, Huajian Xin, Huazuo Gao, Hui Li, Hui Qu, J.~L. Cai, Jian Liang, Jianzhong Guo, Jiaqi Ni, Jiashi Li, Jin Chen, Jingyang Yuan, Junjie Qiu, Junxiao Song, Kai Dong, Kaige Gao, Kang Guan, Lean Wang, Lecong Zhang, Lei Xu, Leyi Xia, Liang Zhao, Liyue Zhang, Meng Li, Miaojun Wang, Mingchuan Zhang, Minghua Zhang, Minghui Tang, Mingming Li, Ning Tian, Panpan Huang, Peiyi Wang, Peng Zhang, Qihao Zhu, Qinyu Chen, Qiushi Du, R.~J. Chen, R.~L. Jin, Ruiqi Ge, Ruizhe Pan, Runxin Xu, Ruyi Chen, S.~S. Li, Shanghao Lu, Shangyan Zhou, Shanhuang Chen, Shaoqing Wu, Shengfeng Ye, Shirong Ma, Shiyu Wang, Shuang Zhou, Shuiping Yu, Shunfeng Zhou, Size Zheng, T.~Wang, Tian Pei, Tian Yuan, Tianyu Sun, W.~L. Xiao, Wangding Zeng, Wei An, Wen Liu, Wenfeng Liang, Wenjun Gao, Wentao Zhang, X.~Q. Li, Xiangyue Jin, Xianzu Wang, Xiao Bi, Xiaodong Liu, Xiaohan Wang, Xiaojin Shen, Xiaokang Chen, Xiaosha Chen, Xiaotao Nie, Xiaowen Sun, Xiaoxiang Wang, Xin Liu, Xin Xie, Xingkai Yu, Xinnan Song, Xinyi Zhou, Xinyu Yang, Xuan Lu, Xuecheng Su, Y.~Wu, Y.~K. Li, Y.~X. Wei, Y.~X. Zhu, Yanhong Xu, Yanping Huang, Yao Li, Yao Zhao, Yaofeng Sun, Yaohui Li, Yaohui Wang, Yi~Zheng, Yichao Zhang, Yiliang Xiong, Yilong Zhao, Ying He, Ying Tang, Yishi Piao, Yixin Dong, Yixuan Tan, Yiyuan Liu, Yongji Wang, Yongqiang Guo, Yuchen Zhu, Yuduan Wang, Yuheng Zou, Yukun Zha, Yunxian Ma, Yuting Yan, Yuxiang You, Yuxuan Liu, Z.~Z. Ren, Zehui Ren, Zhangli Sha, Zhe Fu, Zhen Huang, Zhen Zhang, Zhenda Xie, Zhewen Hao, Zhihong Shao, Zhiniu Wen, Zhipeng Xu, Zhongyu Zhang, Zhuoshu Li, Zihan Wang, Zihui Gu, Zilin Li, and Ziwei Xie.
\newblock Deepseek-v2: A strong, economical, and efficient mixture-of-experts language model, 2024.

\bibitem{TGI}
Hugging Face.
\newblock {Text Generation Inference}.
\newblock \url{https://huggingface.co/text-generation-inference}, 2024.

\bibitem{fei2024empathyear}
Hao Fei, Han Zhang, Bin Wang, Lizi Liao, Qian Liu, and Erik Cambria.
\newblock Empathyear: An open-source avatar multimodal empathetic chatbot, 2024.

\bibitem{FlashInfer}
flashInfer.ai.
\newblock flashinfer.
\newblock \url{https://github.com/flashinfer-ai/flashinfer}, 2023.

\bibitem{gao2024costefficientlargelanguagemodel}
Bin Gao, Zhuomin He, Puru Sharma, Qingxuan Kang, Djordje Jevdjic, Junbo Deng, Xingkun Yang, Zhou Yu, and Pengfei Zuo.
\newblock Cost-efficient large language model serving for multi-turn conversations with cachedattention, 2024.

\bibitem{gim2024promptcache}
In~Gim, Guojun Chen, Seung seob Lee, Nikhil Sarda, Anurag Khandelwal, and Lin Zhong.
\newblock Prompt cache: Modular attention reuse for low-latency inference, 2024.

\bibitem{guan2024fractal}
Yue Guan, Changming Yu, Yangjie Zhou, Jingwen Leng, Chao Li, and Minyi Guo.
\newblock Fractal: Joint multi-level sparse pattern tuning of accuracy and performance for dnn pruning.
\newblock In {\em Proceedings of the 29th ACM International Conference on Architectural Support for Programming Languages and Operating Systems, Volume 3}, ASPLOS '24, page 416–430, New York, NY, USA, 2024. Association for Computing Machinery.

\bibitem{guo2020accelerating}
Cong Guo, Bo~Yang Hsueh, Jingwen Leng, Yuxian Qiu, Yue Guan, Zehuan Wang, Xiaoying Jia, Xipeng Li, Minyi Guo, and Yuhao Zhu.
\newblock Accelerating sparse dnn models without hardware-support via tile-wise sparsity.
\newblock In {\em SC20: International Conference for High Performance Computing, Networking, Storage and Analysis}, pages 1--15. IEEE, 2020.

\bibitem{guo2022squant}
Cong Guo, Yuxian Qiu, Jingwen Leng, Xiaotian Gao, Chen Zhang, Yunxin Liu, Fan Yang, Yuhao Zhu, and Minyi Guo.
\newblock {SQ}uant: On-the-fly data-free quantization via diagonal hessian approximation.
\newblock In {\em International Conference on Learning Representations}, 2022.

\bibitem{Olive}
Cong Guo, Jiaming Tang, Weiming Hu, Jingwen Leng, Chen Zhang, Fan Yang, Yunxin Liu, Minyi Guo, and Yuhao Zhu.
\newblock {OliVe: Accelerating Large Language Models via Hardware-friendly Outlier-Victim Pair Quantization}.
\newblock In {\em Proceedings of the 50th Annual International Symposium on Computer Architecture (ISCA)}. {ACM}, 2023.

\bibitem{guo2024accelerating}
Cong Guo, Fengchen Xue, Jingwen Leng, Yuxian Qiu, Yue Guan, Weihao Cui, Quan Chen, and Minyi Guo.
\newblock Accelerating sparse dnns based on tiled gemm.
\newblock {\em IEEE Transactions on Computers}, 2024.

\bibitem{guo2022ant}
Cong Guo, Chen Zhang, Jingwen Leng, Zihan Liu, Fan Yang, Yunxin Liu, Minyi Guo, and Yuhao Zhu.
\newblock Ant: Exploiting adaptive numerical data type for low-bit deep neural network quantization.
\newblock In {\em 2022 55th IEEE/ACM International Symposium on Microarchitecture (MICRO)}, pages 1414--1433. IEEE, 2022.

\bibitem{guo2024gmlake}
Cong Guo, Rui Zhang, Jiale Xu, Jingwen Leng, Zihan Liu, Ziyu Huang, Minyi Guo, Hao Wu, Shouren Zhao, Junping Zhao, and Ke~Zhang.
\newblock Gmlake: Efficient and transparent gpu memory defragmentation for large-scale dnn training with virtual memory stitching, 2024.

\bibitem{han2015deep}
Song Han, Huizi Mao, and William~J Dally.
\newblock Deep compression: Compressing deep neural networks with pruning, trained quantization and huffman coding.
\newblock {\em arXiv preprint arXiv:1510.00149}, 2015.

\bibitem{hu2024tetri}
Cunchen Hu, Heyang Huang, Liangliang Xu, Xusheng Chen, Jiang Xu, Shuang Chen, Hao Feng, Chenxi Wang, Sa~Wang, Yungang Bao, Ninghui Sun, and Yizhou Shan.
\newblock Inference without interference: Disaggregate llm inference for mixed downstream workloads, 2024.

\bibitem{LMDeploy}
InternLM.
\newblock {LMDeploy: Toolkit for Compressing, Deploying, and Serving Large Language Models}.
\newblock \url{https://github.com/InternLM/lmdeploy}, 2024.

\bibitem{jiang2024mixtral}
Albert~Q. Jiang, Alexandre Sablayrolles, Antoine Roux, Arthur Mensch, Blanche Savary, Chris Bamford, Devendra~Singh Chaplot, Diego de~las Casas, Emma~Bou Hanna, Florian Bressand, Gianna Lengyel, Guillaume Bour, Guillaume Lample, Lélio~Renard Lavaud, Lucile Saulnier, Marie-Anne Lachaux, Pierre Stock, Sandeep Subramanian, Sophia Yang, Szymon Antoniak, Teven~Le Scao, Théophile Gervet, Thibaut Lavril, Thomas Wang, Timothée Lacroix, and William~El Sayed.
\newblock Mixtral of experts, 2024.

\bibitem{kwon2023efficient}
Woosuk Kwon, Zhuohan Li, Siyuan Zhuang, Ying Sheng, Lianmin Zheng, Cody~Hao Yu, Joseph~E. Gonzalez, Hao Zhang, and Ion Stoica.
\newblock Efficient memory management for large language model serving with pagedattention, 2023.

\bibitem{lin2024infinitellm}
Bin Lin, Tao Peng, Chen Zhang, Minmin Sun, Lanbo Li, Hanyu Zhao, Wencong Xiao, Qi~Xu, Xiafei Qiu, Shen Li, Zhigang Ji, Yong Li, and Wei Lin.
\newblock Infinite-llm: Efficient llm service for long context with distattention and distributed kvcache, 2024.

\bibitem{AWQ}
Ji~Lin, Jiaming Tang, Haotian Tang, Shang Yang, Xingyu Dang, and Song Han.
\newblock {AWQ:} activation-aware weight quantization for {LLM} compression and acceleration.
\newblock {\em CoRR}, abs/2306.00978, 2023.

\bibitem{liu2023dejavucontextualsparsity}
Zichang Liu, Jue Wang, Tri Dao, Tianyi Zhou, Binhang Yuan, Zhao Song, Anshumali Shrivastava, Ce~Zhang, Yuandong Tian, Christopher Re, and Beidi Chen.
\newblock Deja vu: Contextual sparsity for efficient llms at inference time, 2023.

\bibitem{mekić2024enhancing}
Edis Mekić, Mihailo Jovanović, Kristijan Kuk, Bojan Prlinčević, and Ana Savić.
\newblock Enhancing educational efficiency: Generative ai chatbots and devops in education 4.0, 2024.

\bibitem{min2023recent}
Bonan Min, Hayley Ross, Elior Sulem, Amir Pouran~Ben Veyseh, Thien~Huu Nguyen, Oscar Sainz, Eneko Agirre, Ilana Heintz, and Dan Roth.
\newblock Recent advances in natural language processing via large pre-trained language models: A survey.
\newblock {\em ACM Computing Surveys}, 56(2):1--40, 2023.

\bibitem{uvm6}
NVIDIA.
\newblock Unified memory in cuda 6, 2013.

\bibitem{pascalgpu}
NVIDIA.
\newblock Pascal architecturec gpu, 2016.

\bibitem{TensorRT}
NVIDIA.
\newblock Nvidia tensorrt.
\newblock \url{https://developer.nvidia.com/tensorrt}, 2023.

\bibitem{GPT4}
OpenAI.
\newblock {GPT-4} technical report.
\newblock {\em CoRR}, abs/2303.08774, 2023.

\bibitem{vLLM042}
OpenLLMAI.
\newblock {vLLM v0.4.2}.
\newblock \url{https://github.com/vllm-project/vllm/releases/tag/v0.4.2}, 2024.

\bibitem{patel2024splitwise}
Pratyush Patel, Esha Choukse, Chaojie Zhang, Aashaka Shah, Íñigo Goiri, Saeed Maleki, and Ricardo Bianchini.
\newblock Splitwise: Efficient generative llm inference using phase splitting, 2024.

\bibitem{pope2022efficiently}
Reiner Pope, Sholto Douglas, Aakanksha Chowdhery, Jacob Devlin, James Bradbury, Anselm Levskaya, Jonathan Heek, Kefan Xiao, Shivani Agrawal, and Jeff Dean.
\newblock Efficiently scaling transformer inference, 2022.

\bibitem{prabhu2024vattention}
Ramya Prabhu, Ajay Nayak, Jayashree Mohan, Ramachandran Ramjee, and Ashish Panwar.
\newblock vattention: Dynamic memory management for serving llms without pagedattention, 2024.

\bibitem{vDNN}
Minsoo Rhu, Natalia Gimelshein, Jason Clemons, Arslan Zulfiqar, and Stephen~W. Keckler.
\newblock vdnn: virtualized deep neural networks for scalable, memory-efficient neural network design.
\newblock In {\em The 49th Annual IEEE/ACM International Symposium on Microarchitecture}, MICRO-49. IEEE Press, 2016.

\bibitem{shazeer2019fast}
Noam Shazeer.
\newblock Fast transformer decoding: One write-head is all you need, 2019.

\bibitem{sun2024llumnix}
Biao Sun, Ziming Huang, Hanyu Zhao, Wencong Xiao, Xinyi Zhang, Yong Li, and Wei Lin.
\newblock Llumnix: Dynamic scheduling for large language model serving, 2024.

\bibitem{torres2024evaluation}
L.~A. Torres, Carlos J.~Barrios H, and Yves Denneulin.
\newblock Evaluation of computational and energy performance in matrix multiplication algorithms on cpu and gpu using mkl, cublas and sycl, 2024.

\bibitem{touvron2023llama}
Hugo Touvron, Thibaut Lavril, Gautier Izacard, Xavier Martinet, Marie-Anne Lachaux, Timothée Lacroix, Baptiste Rozière, Naman Goyal, Eric Hambro, Faisal Azhar, Aurelien Rodriguez, Armand Joulin, Edouard Grave, and Guillaume Lample.
\newblock Llama: Open and efficient foundation language models, 2023.

\bibitem{touvron2023llama2}
Hugo Touvron, Louis Martin, Kevin Stone, Peter Albert, Amjad Almahairi, Yasmine Babaei, Nikolay Bashlykov, Soumya Batra, Prajjwal Bhargava, Shruti Bhosale, Dan Bikel, Lukas Blecher, Cristian~Canton Ferrer, Moya Chen, Guillem Cucurull, David Esiobu, Jude Fernandes, Jeremy Fu, Wenyin Fu, Brian Fuller, Cynthia Gao, Vedanuj Goswami, Naman Goyal, Anthony Hartshorn, Saghar Hosseini, Rui Hou, Hakan Inan, Marcin Kardas, Viktor Kerkez, Madian Khabsa, Isabel Kloumann, Artem Korenev, Punit~Singh Koura, Marie-Anne Lachaux, Thibaut Lavril, Jenya Lee, Diana Liskovich, Yinghai Lu, Yuning Mao, Xavier Martinet, Todor Mihaylov, Pushkar Mishra, Igor Molybog, Yixin Nie, Andrew Poulton, Jeremy Reizenstein, Rashi Rungta, Kalyan Saladi, Alan Schelten, Ruan Silva, Eric~Michael Smith, Ranjan Subramanian, Xiaoqing~Ellen Tan, Binh Tang, Ross Taylor, Adina Williams, Jian~Xiang Kuan, Puxin Xu, Zheng Yan, Iliyan Zarov, Yuchen Zhang, Angela Fan, Melanie Kambadur, Sharan Narang, Aurelien Rodriguez, Robert Stojnic, Sergey Edunov, and Thomas Scialom.
\newblock Llama 2: Open foundation and fine-tuned chat models, 2023.

\bibitem{vaswani2023attention}
Ashish Vaswani, Noam Shazeer, Niki Parmar, Jakob Uszkoreit, Llion Jones, Aidan~N. Gomez, Lukasz Kaiser, and Illia Polosukhin.
\newblock Attention is all you need, 2023.

\bibitem{vllm2023apc}
vLLM.
\newblock Automatic prefix caching.
\newblock \url{https://docs.vllm.ai/en/latest/automatic_prefix_caching/apc.html}.

\bibitem{wang2021dual}
Yang Wang, Chen Zhang, Zhiqiang Xie, Cong Guo, Yunxin Liu, and Jingwen Leng.
\newblock Dual-side sparse tensor core.
\newblock In {\em 2021 ACM/IEEE 48th Annual International Symposium on Computer Architecture (ISCA)}, pages 1083--1095. IEEE, 2021.

\bibitem{vLLM}
Kwon Woosuk, Li~Zhuohan, Zhuang Siyuan, Sheng Ying, Zheng Lianmin, Yu~Cody, Gonzalez Joey, Zhang Hao, and Stoica Ion.
\newblock vllm: Easy, fast, and cheap llm serving with pagedattention.
\newblock \url{https://vllm.ai/}, 2023.

\bibitem{xiao2023smoothquant}
Guangxuan Xiao, Ji~Lin, Mickael Seznec, Hao Wu, Julien Demouth, and Song Han.
\newblock {S}mooth{Q}uant: Accurate and efficient post-training quantization for large language models.
\newblock In {\em Proceedings of the 40th International Conference on Machine Learning}, 2023.

\bibitem{280922}
Gyeong-In Yu, Joo~Seong Jeong, Geon-Woo Kim, Soojeong Kim, and Byung-Gon Chun.
\newblock Orca: A distributed serving system for {Transformer-Based} generative models.
\newblock In {\em 16th USENIX Symposium on Operating Systems Design and Implementation (OSDI 22)}, pages 521--538, Carlsbad, CA, July 2022. USENIX Association.

\bibitem{yuan2024lveval}
Tao Yuan, Xuefei Ning, Dong Zhou, Zhijie Yang, Shiyao Li, Minghui Zhuang, Zheyue Tan, Zhuyu Yao, Dahua Lin, Boxun Li, Guohao Dai, Shengen Yan, and Yu~Wang.
\newblock Lv-eval: A balanced long-context benchmark with 5 length levels up to 256k, 2024.

\bibitem{zhang2022opt}
Susan Zhang, Stephen Roller, Naman Goyal, Mikel Artetxe, Moya Chen, Shuohui Chen, Christopher Dewan, Mona Diab, Xian Li, Xi~Victoria Lin, Todor Mihaylov, Myle Ott, Sam Shleifer, Kurt Shuster, et~al.
\newblock Opt: Open pre-trained transformer language models, 2022.

\bibitem{zhao2024atom}
Yilong Zhao, Chien-Yu Lin, Kan Zhu, Zihao Ye, Lequn Chen, Size Zheng, Luis Ceze, Arvind Krishnamurthy, Tianqi Chen, and Baris Kasikci.
\newblock Atom: Low-bit quantization for efficient and accurate llm serving.
\newblock In P.~Gibbons, G.~Pekhimenko, and C.~De Sa, editors, {\em Proceedings of Machine Learning and Systems}, volume~6, pages 196--209, 2024.

\bibitem{zheng2024sglang}
Lianmin Zheng, Liangsheng Yin, Zhiqiang Xie, Chuyue Sun, Jeff Huang, Cody~Hao Yu, Shiyi Cao, Christos Kozyrakis, Ion Stoica, Joseph~E. Gonzalez, Clark Barrett, and Ying Sheng.
\newblock Sglang: Efficient execution of structured language model programs, 2024.

\bibitem{zhong2024distserve}
Yinmin Zhong, Shengyu Liu, Junda Chen, Jianbo Hu, Yibo Zhu, Xuanzhe Liu, Xin Jin, and Hao Zhang.
\newblock Distserve: Disaggregating prefill and decoding for goodput-optimized large language model serving, 2024.

\end{thebibliography}
